\documentclass[one column]{aastex62}

\usepackage{subfigure}
\usepackage{graphicx}

\begin{document}
\title{The roles of individual advanced LIGO detectors on revealing the neutron star properties}
\author{Ming-zhe Han}
\author{Jin-liang Jiang}
\author{Shao-peng Tang}
\author{Yin-jie Li}
\author{Yi-zhong Fan}
\affil{Key Laboratory of Dark Matter and Space Astronomy, Purple Mountain Observatory, Chinese Academy of Sciences, Nanjing, 210033, People's Republic of China.}
\affil{School of Astronomy and Space Science, University of Science and Technology of China, Hefei, Anhui 230026, People's Republic of China.}
\email{Corresponding authors: jiangjl@pmo.ac.cn (JLJ) and yzfan@pmo.ac.cn (YZF)}

\begin{abstract}
In this work we re-analyze the data of GW170817, the first binary neutron star (BNS) merger event, in two ways, including the parameterized equation of state (EoS) method and gravitational wave (GW) parameter estimation analysis. Three sets of data, including the combined LIGO/Virgo (HLV) data, the LIGO-Hanford (H1) data and the LIGO-Livingston (L1) data, have been analyzed, respectively. We confirm the bimodal probability distribution of the tidal deformability of GW170817 with the HLV data and resolve the different contributions of H1 and L1 detectors. Our simulation reveals a tendency of ``overestimating" the tidal parameter with the decreasing signal-to-noise ratio (SNR). Such a tendency is likely caused by the dependence of the result on the prior (which is not well understood and has been assumed to be widely-distributed) in the low SNR case. Though this effect is interesting, the slightly higher SNR of H1 than L1 at frequencies above $\sim 100$ Hz is most-likely not the main reason for the lower tidal parameter region favored by H1. In view of the large fluctuation of the expected tide parameter in the case of low SNR, the different probability distributions favored by the L1 and H1 data of GW170817 are reasonable. We have also explored the dependence of expected difference of SNRs of H1 and L1 detectors on the source locations in the O3 and full sensitivity runs. Such maps are helpful in evaluating the relative powers of individual detectors on measuring the tidal deformabilities for the new double neutron star merger events.
\end{abstract}
%%%%%%%%%%%%%%%%%%%%%%%%%% Introduction %%%%%%%%%%%%%%%%%%%%%%%%%%
\section{Introduction}
\label{sec:Intro}
The gravitational wave signal of the first binary neutron star (BNS) merger event GW170817 was discovered by the LIGO-Virgo detector network on August 17, 2017 \citep{2017PhRvL.119p1101A}. It is a milestone for the studies of neutron star (NS), in particular the investigations on the properties of matters at extreme densities \citep[see][for reviews]{2012ARNPS..62..485L, 2016PhR...621..127L, 2016ARA&A..54..401O, 2017RvMP...89a5007O}. Together with some reasonable
assumptions/EoS-independent relationships, the tidal deformities/radii of the two NSs powering GW170817 have
been inferred and the EoS of NSs has been constrained \citep[e.g.,][]{2018PhRvL.121p1101A, 2019PhRvX...9a1001A, 2018PhRvL.120q2703A, 2018PhRvL.121i1102D, 2018PhRvL.120q2702F, 2018ApJ...868L..22L, 2018PhRvL.121f2701L, 2019arXiv190205502L, 2019ApJ...885...39J, 2019arXiv191108107T}.

For non-spinning neutron stars, the degree of deformation is described by the dimensionless tidal deformability parameter, $\Lambda=(2/3)k_{2}[(c^2/G)(R/m)]^5$, where $k_{2}$ is the tidal Love number \citep{2008ApJ...677.1216H, 2008PhRvD..77b1502F, 2009PhRvD..80h4035D, 2009PhRvD..80h4018B}, $c$ is the speed of light in vacuum, $G$ is the gravitational constant, and $m$ and $R$ are the mass and radius of NS, respectively. The combined tidal parameter $\tilde{\Lambda}$ \citep{2014PhRvD..89j3012W} is given by
$\tilde{\Lambda}=16[(m_{1}+12m_{2})m_{1}^4\Lambda_1+(m_{2}+12m_{1})m_{2}^4\Lambda_2]/13{(m_{1}+m_{2})^5}$,
where $m_{1, 2}$ and $\Lambda_{1, 2}$ are the masses and tidal deformabilities of each NS. Interestingly, the data analysis of GW170817 result in a bimodal probability distribution of $\tilde{\Lambda}$ \citep[e.g.][]{2018PhRvL.121p1101A, 2019PhRvX...9a1001A, 2019ApJ...885...39J}. \citet{2018arXiv181206100N} analyzed the data of the LIGO-Hanford (H1) and the LIGO-Livingston (L1), respectively. They found out that the L1 data yield a higher $\tilde{\Lambda}$ while the H1 data favor a small one, and these two separated peaks give rise to the bimodal probability distribution revealed in the combined HLV data analysis.
Further studies are needed to reveal the underlying reasons. Note that in both the second observing run (O2) and the third observing run (O3), the Advanced Virgo (V1) detector is not as sensitive as the two Advanced LIGO detectors, therefore its role on constraining $\tilde{\Lambda}$ is less important and we do not take into account it in most analysis.

In the first half year of O3 of LIGO/Virgo network\footnote{\url{https://gracedb.ligo.org/superevents/public/O3/}}, there were 33 candidates and three of them are BNS merger candidates ($p_{\rm BNS}>50\%$).  All these three BNS candidates, including the S190425z\footnote{\url{https://gracedb.ligo.org/superevents/S190425z/view/}}, S190901ap\footnote{\url{https://gracedb.ligo.org/superevents/S190901ap/view/}} and S190910h\footnote{\url{https://gracedb.ligo.org/superevents/S190910h/view/}}, were only detected by L1 (except for V1). Therefore, their $\tilde{\Lambda}$ have to be constrained essentially with the absent of H1 and it is thus interesting to investigate whether such constraints are biased or not.
For such a purpose, again we need a reasonable understanding of the different roles of the L1 data and H1 data on bounding $\tilde{\Lambda}$ of GW170817.

Inspired by the above concerns, in this work we re-analyze the data of GW170817. We then explore the underlying reasons for the difference of the L1 and H1 data on bounding $\tilde{\Lambda}$ and finally evaluate the potential ``bias" on inferring $\tilde{\Lambda}$ for the three BNS candidate events found in the first half year of O3 run. This work is organized as follows. In section~\ref{sec:Method} we describe the theoretical and data analysis methods. In section~\ref{sec:Results} we present our results, including the parameter estimation results of GW analysis/parameterized EoS method, and the study of the ``difference" of the reconstructed $\tilde{\Lambda}$ between H1 and L1. Finally we summarize our work.

%%%%%%%%%%%%%%%%%%%%%%%%%%%%%%%%%%% Method %%%%%%%%%%%%%%%%%%%%%%%%%%%%%%%%%%%%%%%%%%%
\section{Methods}
\label{sec:Method}
\subsection{Bayesian Inference}
To study the properties of BNS systems, we need the posterior probability of parameters $\vec{\lambda}$. For the given data $d(t)$, the posterior probability can be constructed with the Bayes theorem, i.e.,
\begin{equation}
\label{eq:bayes}
p(\vec{\lambda}|d(t), M)=\frac{\mathcal{L}(d(t)|\vec{\lambda}, M)p(\vec{\lambda}|M)}{p(d(t)|M)},
\end{equation}
where $p(\vec{\lambda}|d(t), M)$ is the posterior probability density function (PDF), $\mathcal{L}(d(t)|\vec{\lambda}, M)$ is the likelihood, $p(\vec{\lambda}|M)$ is the prior PDF, $M$ is the model of data (i.e., the signal and noise models), and $p(d(t)|M)$ is the evidence (which is not important for parameter estimation unless for model selection).

When the gravitational wave reaches the detector, the detector records
\begin{equation}
\label{eq:data}
d(t)=n+R[h(\vec{\lambda})],
\end{equation}
i.e., the data is consisted of detector noise and response of the gravitational wave. If we assume the noise of detector is stationary Gaussian noise, then the probability of obtaining data $d(t)$ with signal $h(\vec{\lambda})$ can be written as
\begin{equation}
\label{likelihood}
\mathcal{L}(d(t)|\vec{\lambda}, M) \approx \exp\{-\frac{1}{2}(d(t)-R[h(\vec{\lambda})]|d(t)-R[h(\vec{\lambda})])\},
\end{equation}
where
\begin{equation}
(a|b)=2\int^{+\infty}_{0}df\frac{a^*(f)b(f)+a(f)b^*(f)}{S_{\rm n}(f)},
\end{equation}
the $S_{\rm n}(f)$ is one-sided power spectral density (PSD). The PSDs we used for our parameter estimation are provided by the LIGO Document Control Center \footnote{\url{https://dcc.ligo.org/LIGO-P1800061/public}} \citep{2019PhRvX...9a1001A}, and the strain data we used are available at the Gravitational Wave Open Science Center \footnote{\url{https://www.gw-openscience.org/events/GW170817/}} \citep{2015JPhCS.610a2021V}.
\subsection{Parameterizing EoS}
Parameterized EoS method \citep{2010PhRvD..82j3011L, 2014ApJ...789..127K, 2019arXiv190205502L, 2019arXiv190404233M, 2016EPJA...52...18S} is very useful to study the properties of dense matters in the core of NS. Two widely-adopted methods are the spectral expansion \citep{2010PhRvD..82j3011L} and the piecewise polytropic expansion \citep{2009PhRvD..79l4032R}. In this work we adopt the second model, i.e., the piecewise polytropic expansion, to parameterize the EoS. The EoS can be expressed as
\begin{equation}
P=K\rho^\Gamma,
\end{equation}
where $P$ denotes the pressure, $\rho$ denotes the mass density, $K$ is constant in each piece of parameterized EoS, and $\Gamma$ is the adiabatic index. We parameterize the EoS with four pressures $\{P_{1}, P_{2}, P_{3}, P_{4}\}$ at corresponding densities, $\{1, 1.85, 3.7, 7.4\}\rho_{\rm sat}$ \citep{2009PhRvD..80j3003O}, where $\rho_{\rm sat}=2.7 \times 10^{14} \rm g/\rm cm^3 $ is the so-called saturation density.

Now we have parameterized the EoS with pressures. Before applying this method we have one more parameter left, the central pseudo enthalpy $h_{\rm c}$. The $h_{\rm c}$ is defined as
\begin{equation}
h_{\rm c} \equiv \int^{p_{\rm c}}_{0} \frac{dp}{\epsilon(p)+p},
\end{equation}
where $p$, $p_{\rm c}$, and $\epsilon$ denotes the pressure, the central pressure of the NS, and the energy density, respectively. Therefore, we can calculate $\{m, R, \Lambda\}$ from the parametrization parameters $\{h_{\rm c}, P_{1}, P_{2}, P_{3}, P_{4}\}$ (see our previous work \cite{2019ApJ...885...39J} based on \cite{2014PhRvD..89f4003L}). Following \cite{2019ApJ...885...39J} and \cite{2019arXiv191108107T}, in this work we take the dimensionless parameters $\{\hat{p}_{1}, \hat{p}_{2}, \hat{p}_{3}, \hat{p}_{4}\}$ instead of $\{P_{1}, P_{2}, P_{3}, P_{4}\}$, where $\hat{p}_{\rm i}=P_{\rm i}/(10^{32+\rm i} ~\rm dyn ~\rm cm^{-2})$.

\subsection{Priors}
For GW analysis, we samples
\begin{equation}
\mathcal{M} = \frac{(m_1m_2)^{3/5}}{(m_1+m_2)^{1/5}}
\end{equation}
and $q=m_2/m_1$ in place of $m_1$ and $m_2$. The mass priors we choose for the analysis are both uniform, and the ranges are $\mathcal{M} \in (1.184, 2.168)\,M_{\odot}$ and $q \in (0.5, 1)$. The dimensionless spins of each component $\chi_{1,2}$ are constrained to be aligned with the orbital-angular momentum and the ranges are $\chi_{1,2} \in (-0.05, 0.05)$. The dimensionless tidal deformability parameters $\Lambda_{1, 2}$ are uniformly spaced in $(0, 5000)$. We fix the sky location of the binary to the right ascension RA$ =197.450374^{\circ}$ and declination Dec$=-23.381495^{\circ}$ \citep{2017ApJ...848L..16S}. Luminosity distance prior is a Gaussian distribution with $ \mu = 40.7 ~ {\rm Mpc}$ \citep{2018ApJ...854L..31C} and the dispersion of $\sigma=2.36 ~ {\rm Mpc}$. Other external parameters' priors are all uniform in their allowed domain.

For parameterized EoS method, we samples in $\{\mathcal{M}, q, \hat{p}_{1}, \hat{p}_{2}, \hat{p}_{3}, \hat{p}_{4} \}$ in place of $\{m_1, m_2, \Lambda_1, \Lambda_2 \}$. The priors of these parameters are all uniform, $\mathcal{M}\in (1.184, 2.168)\,M_{\odot}$, $q\in (0.5, 1)$, $\hat{p}_{1}\in (1.5, 13.5)$, $\hat{p}_{2}\in (0.7, 8.0)$, $\hat{p}_{3}\in (0.6, 7.0)$, and $\hat{p}_{4}\in (0.3, 4.0)$. Additional constraints are applied on the parameterized EoS including \citep[see also][]{2017ApJ...844..156R, 2019ApJ...885...39J}
\begin{itemize}
\item Microscopical stability $P_4 \ge P_3 \ge P_2 \ge P_1$.
\item Causality condition $c_{\rm s}/c \le 1$, where $c_{\rm s}$ is the sound speed.
\item Maximum stable mass of non-rotating limit $M_{\rm TOV} \in (2.06, 2.5)M_{\odot}$ \citep{1998PhRvC..58.1804A, 2016PhR...621..127L, 2018ApJ...858...74M, 2019NatAs.tmp..439C}
\item Adiabatic indexes limit $\Gamma < 7$.
\end{itemize}

We do not consider the calibration errors of the detectors since they do not significantly affect the posterior of masses, spins, or tidal parameters \citep{2019PhRvX...9a1001A}. The waveform we choose is PhenomDNRT, which is based on an aligned-spin point-particle \citep{2016PhRvD..93d4007K, 2016PhRvD..93d4006H} model calibrated to untuned EOB waveforms \citep{2014PhRvD..89f1502T} and NR hybrids \citep{2016PhRvD..93d4007K, 2016PhRvD..93d4006H}.

\begin{figure}
\centering
\subfigure{
\includegraphics[width=0.43\columnwidth]{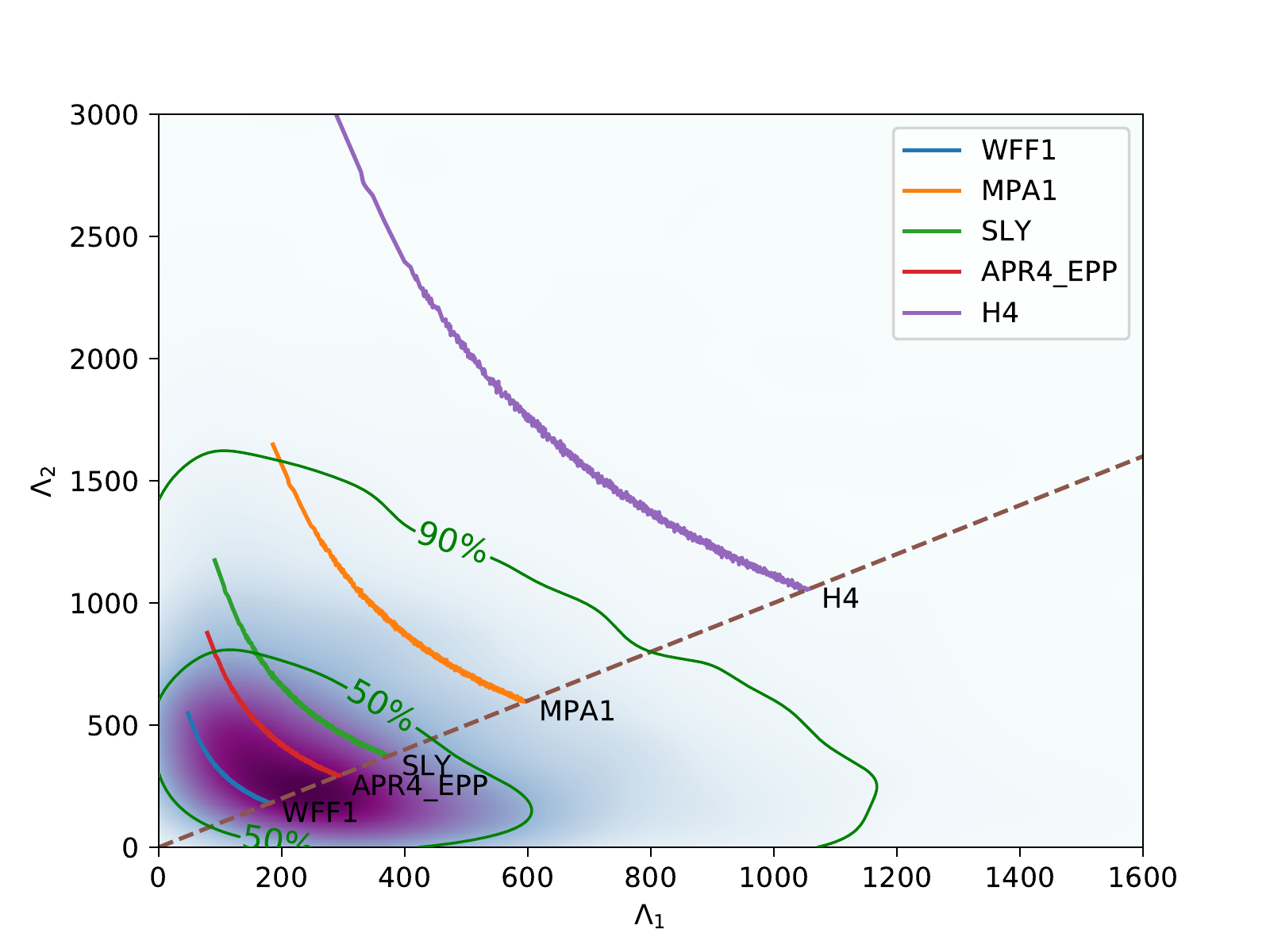}
}
\subfigure{
\includegraphics[width=0.43\columnwidth]{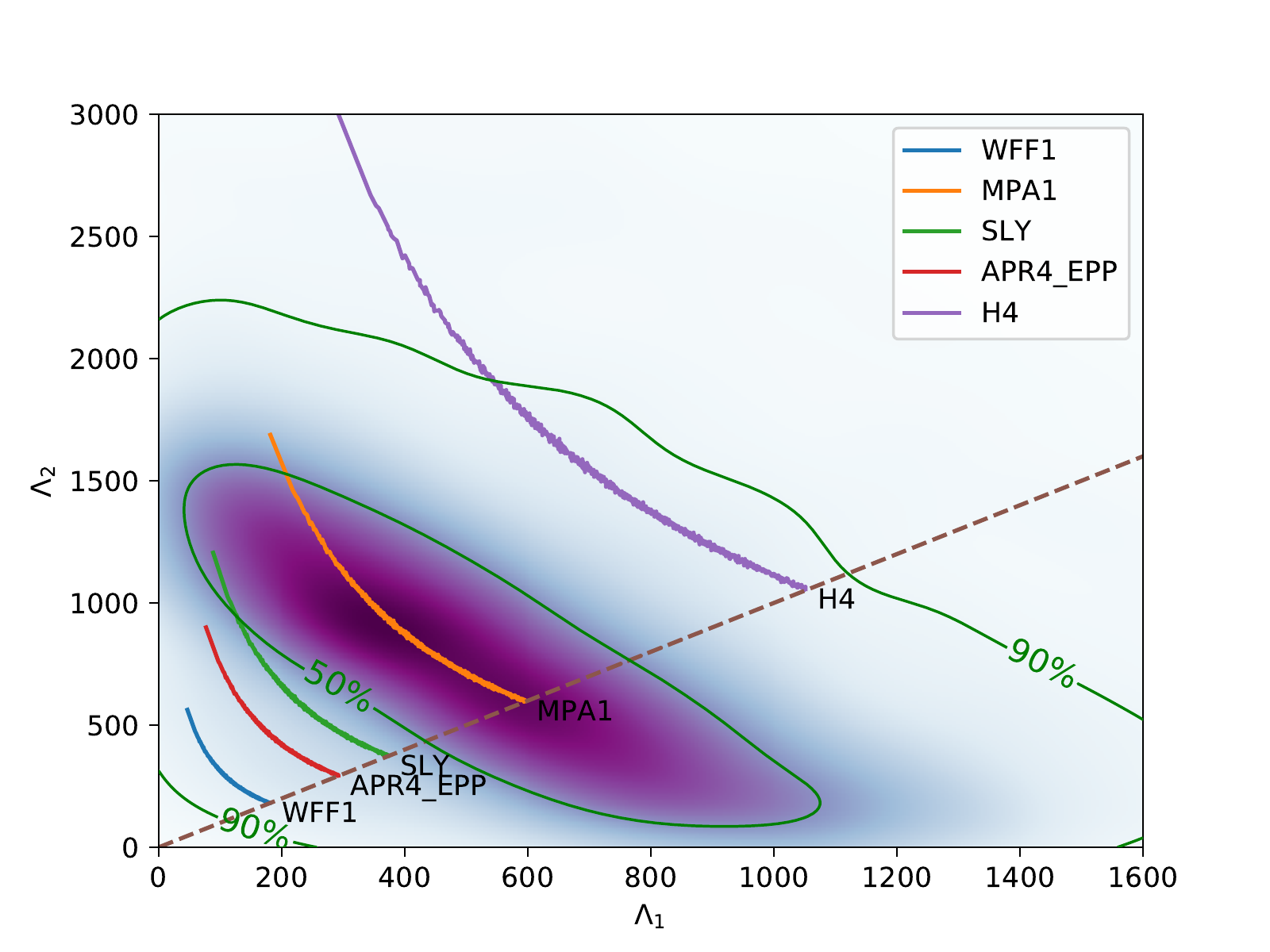}
}
\caption{PDFs of the dimensionless tidal deformability $\Lambda_{1}$ and $\Lambda_{2}$. The left panel shows the result of individual H1 detector and the right panel shows the result of individual L1 detector. The 50\% and 90\% credible regions are shown in green contours. The five colored curves are the tidal parameters calculated using the mass posterior with different EoS models, including H4 \citep{2006PhRvD..73b4021L}, MPA1 \citep{1987PhLB..199..469M}, SLY \citep{2001A&A...380..151D}, APR4\_EPP \citep{1998PhRvC..58.1804A}, and WFF1 \citep{1988PhRvC..38.1010W}. The dashed line is the $\Lambda_{1}=\Lambda_{2}$ boundary.}
\label{fig:lam12}
\end{figure}

We use the Bilby \citep{2019ascl.soft01011A}, a python package based on Bayesian analysis for parameter estimations of gravitational wave data, to estimate the posterior probabilities of the parameters, and the sampler we choose is PyMultiNest \citep{2016ascl.soft06005B}. The likelihood of parameterized EoS method is calculated by PyCBC \citep{2018ascl.soft05030T, 2019PASP..131b4503B}, and the EoS curves in Fig.\ref{fig:lam12} is calculated by LALSimulation package \citep{2018PhRvD..98f3004C, lalsuite}. In the analysis of GW170817, we focus on the following three data sets, including
\begin{itemize}
\item Only the LIGO-Hanford detector data (H1)
\item Only the LIGO-Livingston detector data (L1)
\item The combined Advanced LIGO and Advanced Virgo detector data (HLV)
\end{itemize}

%%%%%%%%%%%%%%%%%%%%%%%%%%%%%%%%%%%% Results %%%%%%%%%%%%%%%%%%%%%%%%%%%%%%%%%%

\section{Results}
\label{sec:Results}
\subsection{Physical parameters of neutron stars: constraints with the data sets of GW170817}
In Fig~\ref{fig:lam12}, we use the posterior of our GW analysis results to plot the density of $\Lambda_{1}$ and $\Lambda_{2}$ with the tidal parameter curves of different EoS models. Evidently, the H1 data (left panel) prefer a more compact NS than the L1 data (right panel). In particular, the EoS model H4 totally lies outside the $90\%$ credible region for the H1 data, which it is not the case for the L1 data. Thus, if we had only H1 (L1) data, the H4 EoS model is strongly-disfavored (still acceptable).

\begin{figure}
\centering
\subfigure{
\includegraphics[width=0.4\columnwidth]{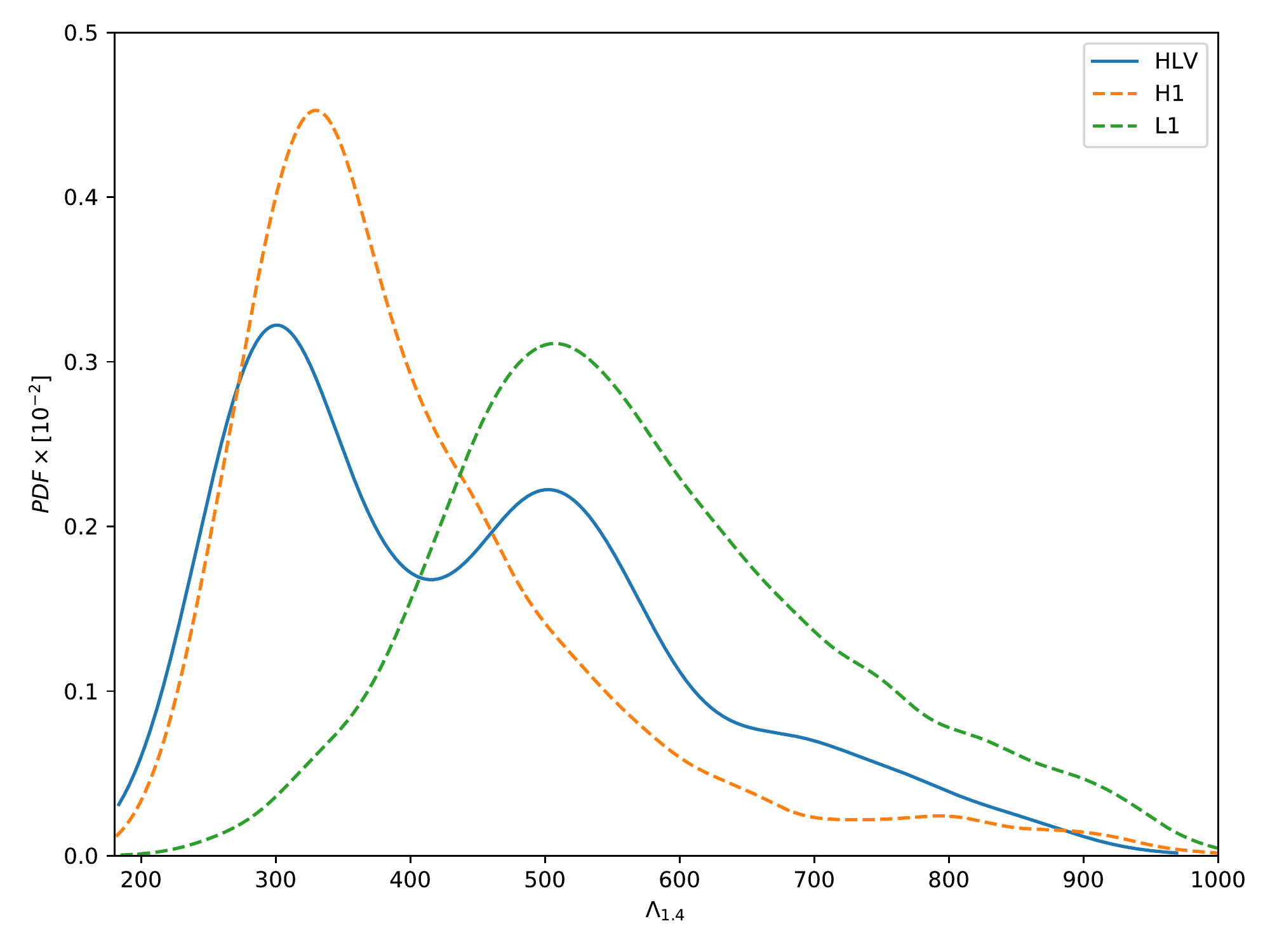}
}
\subfigure{
\includegraphics[width=0.4\columnwidth]{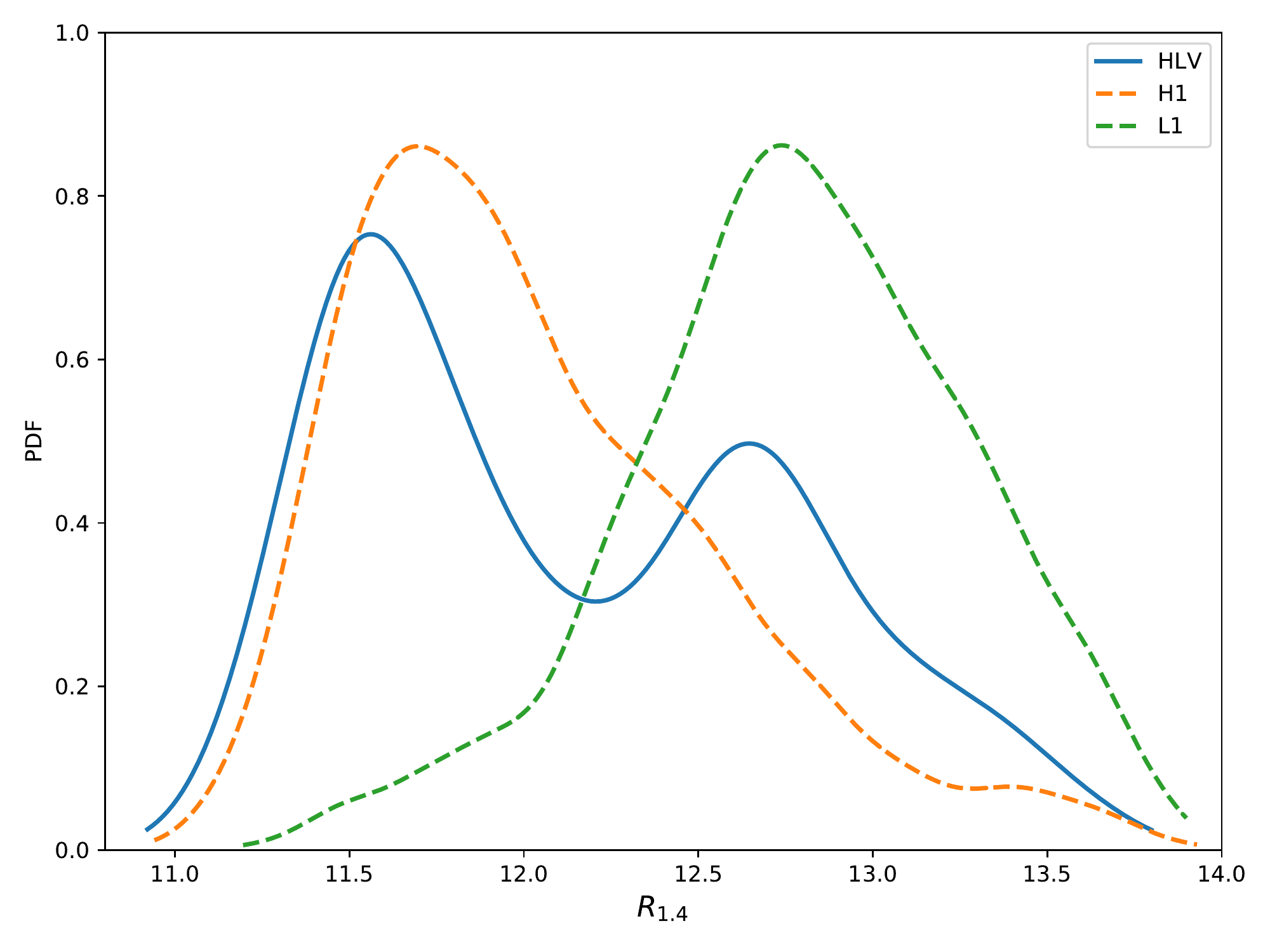}
}
\caption{Canonical global properties of the NSs evaluated from posterior samples of the parameterized EoS analysis. The left panel shows the posterior PDFs of $\Lambda_{1.4}$, and the right panel shows the PDFs of $R_{1.4}$. Blue solid line denotes the combined HLV analysis, and colored dashed lines denote the individual detector analysis.}
\label{fig:LR}
\end{figure}

With a specific equation of state, we can map the property $M=1.4M_{\odot}$ to the properties $\Lambda_{\rm 1.4}$ and $R_{\rm 1.4}$ through solving the TOV equations and the Regge-Wheeler equation. Thus we can get different distributions of $\Lambda_{\rm 1.4}$ and $R_{\rm 1.4}$ if different sets of EoSs are given (as shown in Fig.\ref{fig:LR}). Clearly, the bimodal probability distributions of both $\Lambda_{1.4}$ \citep[see also the case of $\tilde{\Lambda}$ in][]{2018PhRvL.121p1101A, 2019PhRvX...9a1001A, 2018arXiv181206100N} and $R_{1.4}$ found in the joint HLV data analysis are due to the ``separated" peaks of the corresponding probability distributions given by the data of two individual LIGO detectors. Though the contribution of the L1 data is important, the probability distributions of both $\Lambda_{1.4}$ and $R_{1.4}$ are dominated by the H1 data. The investigation of the underlying reason is the focus of the next subsection.

\subsection{On the different performance of L1 and H1 detectors}
\subsubsection{The parameter estimation in the low SNR case}
\begin{figure}
\centering
\includegraphics[width=0.6\columnwidth]{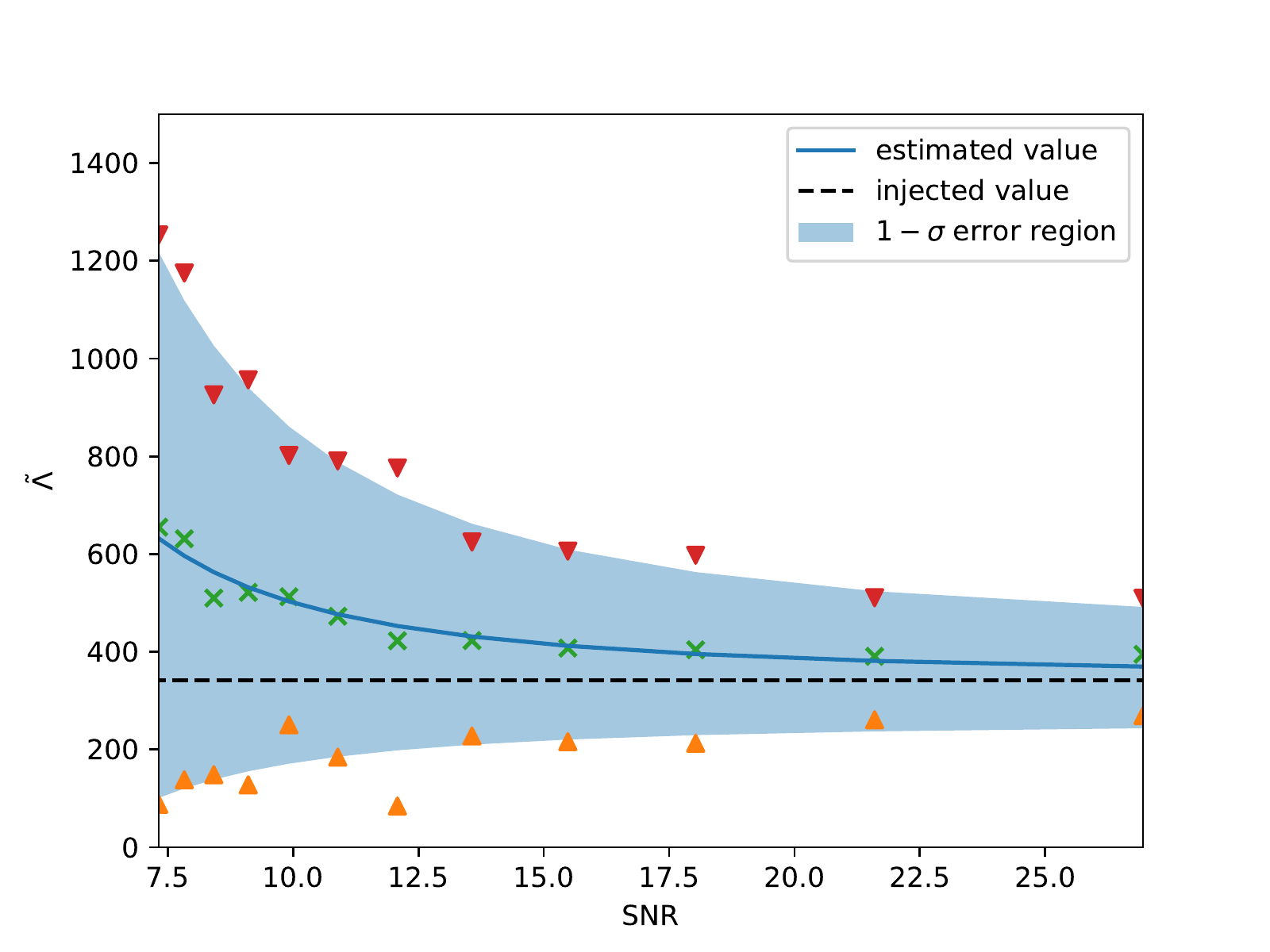}
\caption{The simulated results of $\tilde{\Lambda}$ at different distances with different SNRs, where we only consider the contribution of high frequency data ($>150$ Hz) to the SNR, and all the other parameters are the same as the best-fit parameters of GW170817. At each distance we carry out 40 simulations and take the median value, the solid curve and shadow area are the smoothed median value and 1-sigma uncertainties region, and the horizontal dashed line represents the injected value. The original results are dotted in the figure.}
\label{fig:ld_dis}
\end{figure}{}
Within the framework of Bayesian parameter estimation, the signal-to-noise ratio (SNR) roughly qualifies the difficulty of extracting information from the GW data. Usually, low SNR will lead to less divergence between the inferred posterior and the priors, thus make loose constraints on the parameters. With the decreasing of SNR, the posterior distributions of the high post-Newtonian (PN) parameters that have minor contribution to GW waveforms will approach their prior distributions. Therefore, if we take a naive priors, e.g., $\Lambda \sim U(0,5000)$, it is understandable to obtain a biased larger $\Lambda$ in the parameter estimations.

To test this speculation, we make some simulations as follows. First, we generate a series of ``events" with different SNRs using the best-fit values of the parameters of GW170817 except luminosity distance, which varies between $40-150 \rm Mpc$. For each ``event" at specific distance, we inject the signal with GW parameters and recover them using Bayesian parameter estimation. To avoid any occasional factors and make a robust evaluation of the uncertainties, we carry out 40 simulations at each distance and average over the median values and the corresponding $1\sigma$ uncertainties of $\tilde{\Lambda}$. As shown in Fig.\ref{fig:ld_dis}, roughly the parameter estimation result of the $\tilde{\Lambda}$ decreases as the SNR increases. In other words, a lower SNR tends to yield a larger $\tilde{\Lambda}$, as anticipated. The other fact one should bear in mind is the large fluctuations of the parameter estimation in the case of low SNR.

\subsubsection{The different SNR of the L1 and H1 data}
\begin{figure}
\centering
\subfigure{
\includegraphics[width=0.4\columnwidth]{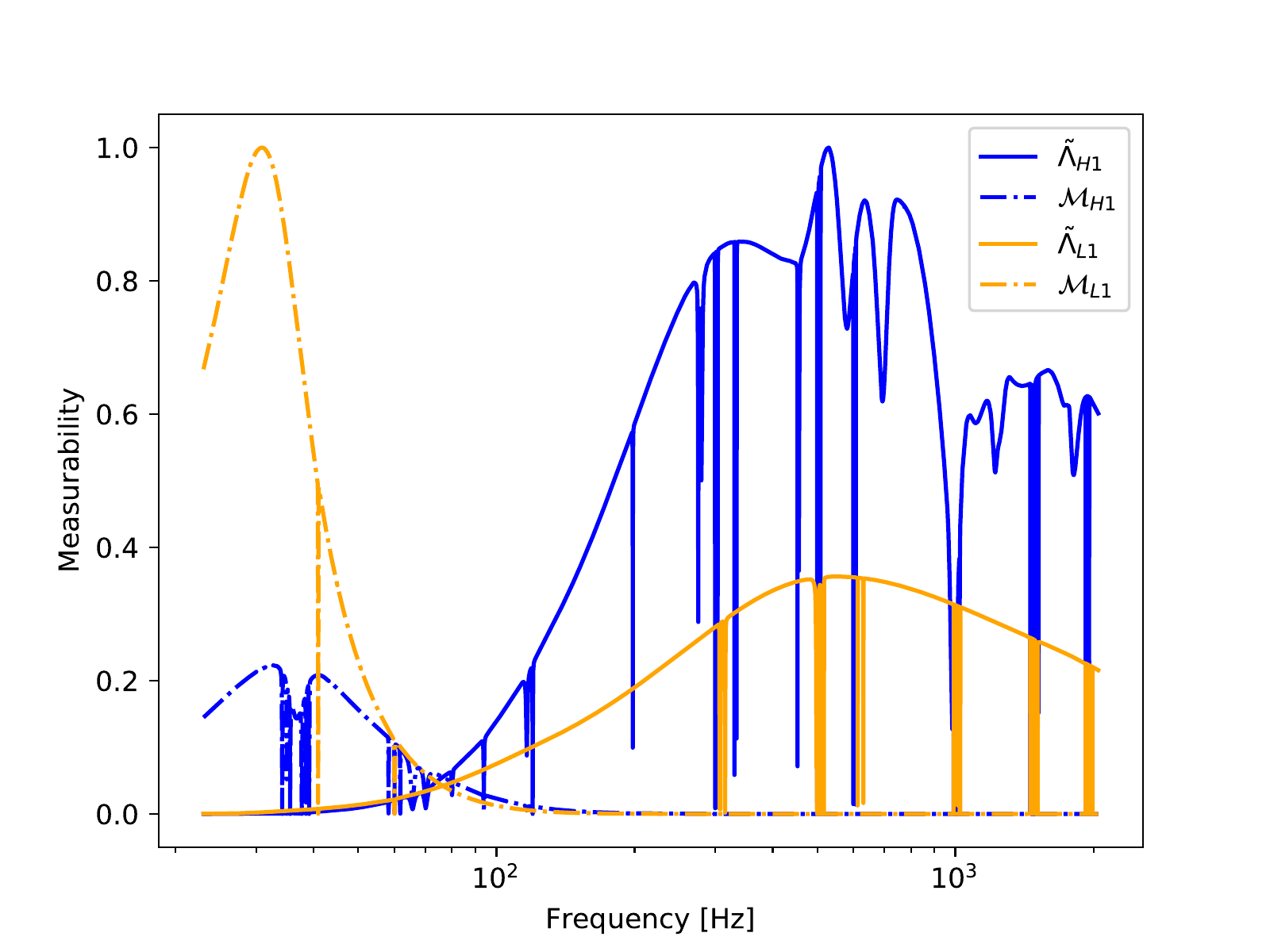}
}
\subfigure{
\includegraphics[width=0.4\columnwidth]{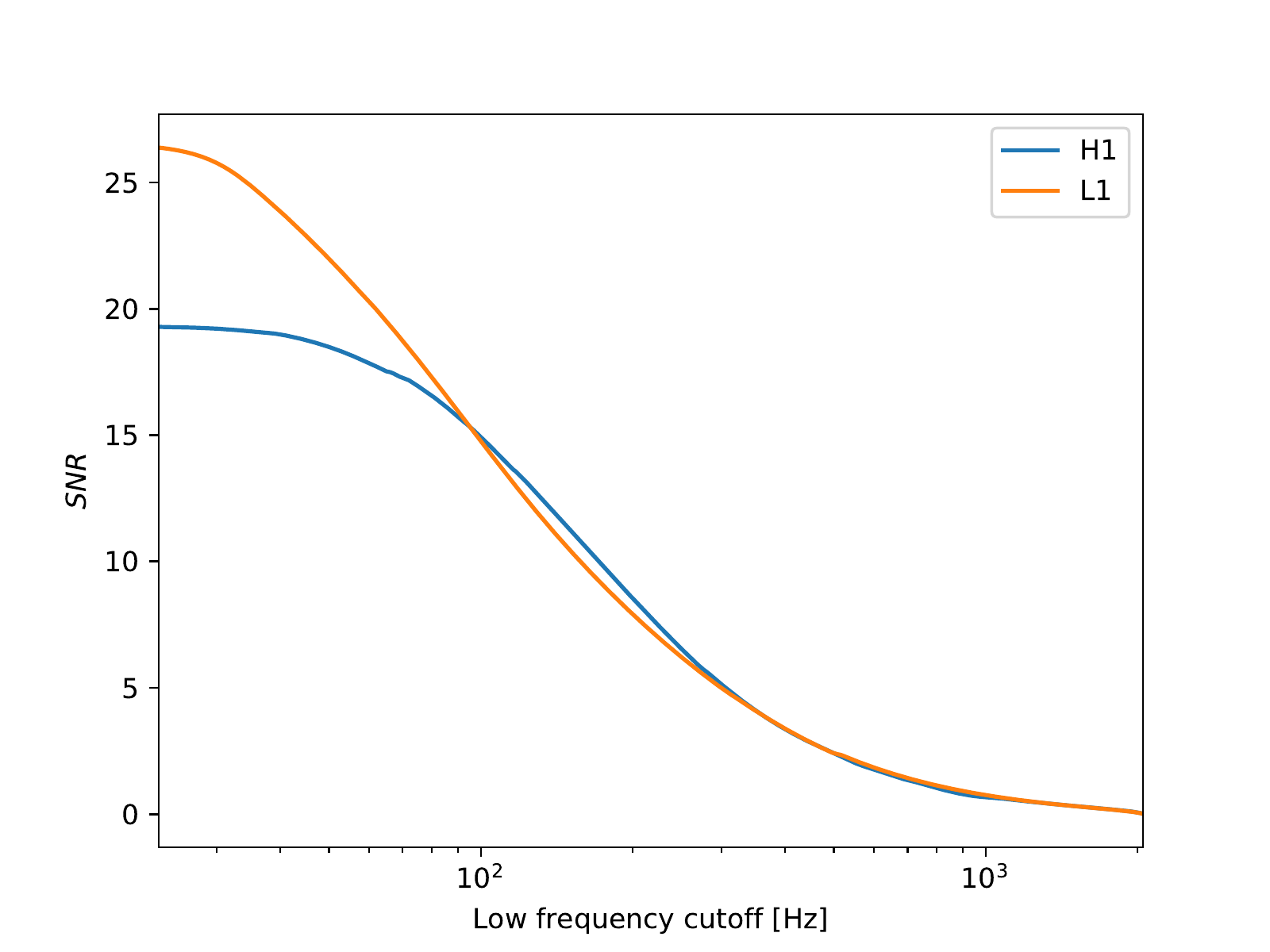}
}
\caption{The left panel shows the normalized measurabilities of $\mathcal{M}$ and $\tilde{\Lambda}$ for H1 and L1. The solid lines denote the $\tilde{\Lambda}$, while the dashed lines denote the $\mathcal{M}$, and the blue, orange lines denote H1, L1, respectively. Note that for the two Advanced LIGO detectors, the measurability of the $\tilde{\Lambda}$ for H1 is better than L1, while for $\mathcal{M}$ it is opposite. The right panel shows the SNRs of the two Advanced LIGO detectors for different low frequency cutoff of the strain data in GW170817.}
\label{fig:170817_freq}
\end{figure}

In the late inspiral phase of GW170817, a glitch took place in the LIGO-Livingston detector \citep{2017PhRvL.119p1101A}, which may have a tremendous impact on the estimation of tidal parameters. The PyCBC gating algorithm \citep{2016CQGra..33u5004U} and {\tt BayesWave} algorithm \citep{2015CQGra..32m5012C, 2015PhRvD..91h4034L} can remove the glitch. \cite{2018PhRvD..98h4016P} has studied the effects of these techniques, and demonstrated with the simulated data that these techniques are able to produce unbiased measurements of the intrinsic parameters.
Moreover, it is possible to directly check whether the low SNR of L1 data ($>100$ Hz) of GW170817 is solely due to the presence of the glitch. For such a purpose, we evaluate the measurabilities of $\mathcal{M}$ and $\tilde{\Lambda}$ in the frequency range $23-2048$ Hz. The measurabilities of $\mathcal{M}$ and $\tilde{\Lambda}$ are associated with the integrand of $I_{-10}$ and $I_{+10}$ \citep[see][]{2012PhRvD..85l3007D}. The left panel of Fig.\ref{fig:170817_freq} shows the measurabilities for the two LIGO detectors. The measurability of $\mathcal{M}$ is mainly governed by the low frequency ($<100$ Hz) data while the measurability of $\tilde{\Lambda}$ is mainly contributed by the high frequency ($>100$ Hz) data. Thus the quality of the data at high frequencies (late inspiral up to the merger) has the major influence on the estimation of $\tilde{\Lambda}$. As shown in the left panel of Fig.\ref{fig:170817_freq}, it is evident that H1 is more sensitive on $\tilde{\Lambda}$ than L1, while it is opposite on $\mathcal{M}$. This suggests a more important role of H1 data on bounding $\tilde{\Lambda}$ of GW170817.

According to the conclusion we got from the measurability (i.e., $\tilde{\Lambda}$ is governed by high frequency data), if we compute the difference between SNRs of individual detectors with higher frequency data, we can use the relation between the parameter estimation result of $\tilde{\Lambda}$ and SNR to figure out whether the discrepancy of $\tilde{\Lambda}$ is caused by the difference of SNR. Thus we compute the difference between the SNRs of the two Advanced LIGO detectors for GW170817 with various low frequency cutoff, and the results are shown in the right panel of Fig.\ref{fig:170817_freq}. With a low frequency cutoff of 23 Hz, the SNR of L1 is larger than that of H1 by $\sim 7$, which is well consistent with the result in GWTC-1 \citep[see TABLE V in][]{2019PhRvX...9c1040A}. If we just consider the high frequency ($>100$ Hz) data, which is significant for the estimation of tidal deformability parameter, the SNR of individual L1 will be lower than H1, and the parameter estimation result of $\tilde{\Lambda}$ will be higher than H1. Though interesting, the slightly higher SNR of H1 than L1 at frequencies above $\sim 100$ Hz is not the main reason for the lower tidal parameter region favored by the H1 data. Due to the large fluctuation of the expected tide parameter in the case of low SNR, the different probability distributions favored by the L1 and H1 data of GW170817 are reasonable.

\subsection{The spatial dependence of SNRs of H1/L1}
\begin{figure}
\centering
\subfigure{
\includegraphics[width=0.45\columnwidth]{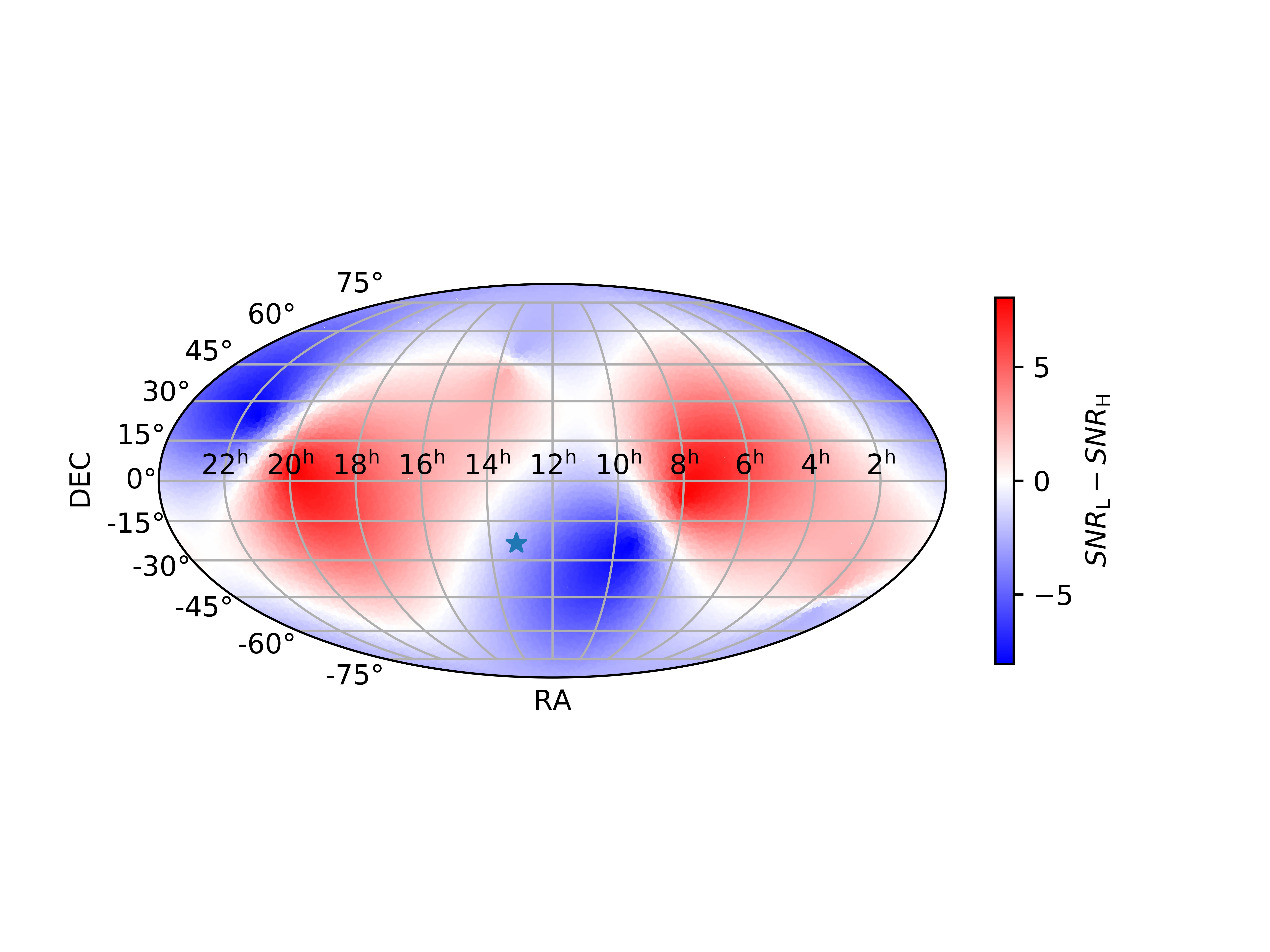}
}
\subfigure{
\includegraphics[width=0.45\columnwidth]{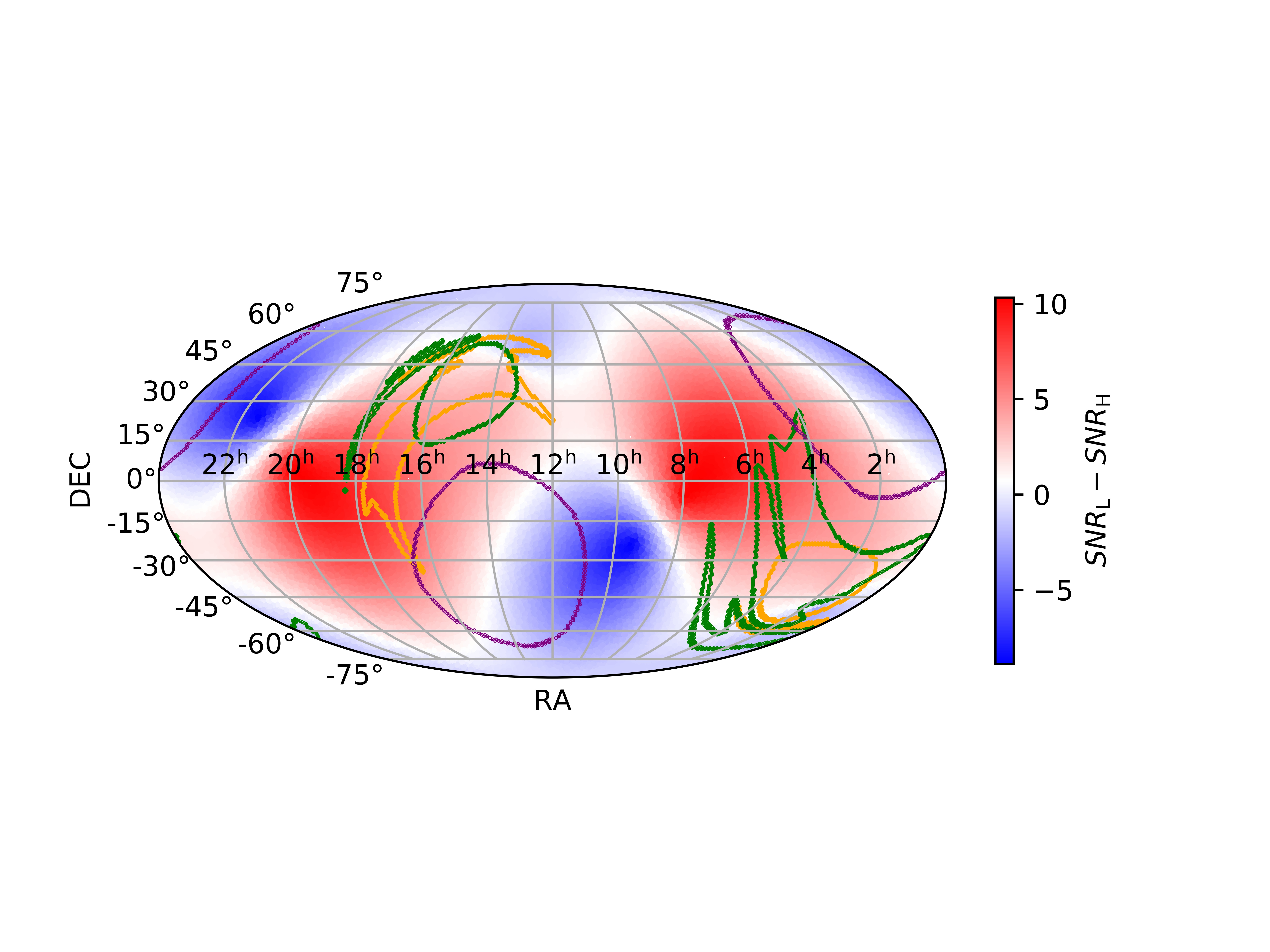}
}
\caption{The difference between the SNRs of the two Advanced LIGO detectors for different sky locations, where all the other parameters are the same as the best-fit parameters of GW170817. The left panel uses the aLIGODesignSensitivity PSD and the right panel uses the O3's PSDs. Both results are for a low frequency cutoff of 150 Hz since we focus on the tidal effects. The red region denotes the SNR of individual H1 is lower than that of L1, while the blue region denotes the SNR of L1 is lower than that of H1. The blue star in the left panel is the localization of GW170817, the orange, green, and purple contours denote the 50\% credible regions of localization of BNS candidate S190425z, S190901ap, and S190910h, respectively.}
\label{fig:snr_sky}
\end{figure}{}
We have shown above that the different favored regions of $\tilde{\Lambda}$ for GW170817 are likely due to the different SNRs of the H1 and L1 data. The different SNRs are partly due to the different sensitivities of the H1 and L1 detectors. Interestingly, as shown below, even with the same sensitivity, usually the two LIGO detectors will yield different SNRs.

We use the aLIGODesignSensitivity PSD \citep{2018LRR....21....3A} for both L1 and H1, and fix the parameters' values to be the best-fit values of the parameters of GW170817 except the sky location, namely right ascension (RA) and declination (DEC), to make the simulated injection signals. Then we compute the SNRs for each simulated injection signal, and the result is shown in the left panel of Fig.\ref{fig:snr_sky}. The pattern implies the relation between the differences of SNRs in H1/L1 and the sky location of the source. For a fixed distance, $40.7 ~\rm Mpc$, the relative SNR difference varies with the sky location. This is reasonable since the two detectors have different antenna response functions \citep{2012PhRvD..85l2006A}. Besides, GW170817 locates in the blue region, which confirms our previous conclusion. Furthermore, in the right panel of Fig.\ref{fig:snr_sky}, we change the PSDs to the O3's PSDs \footnote{\url{https://dcc.ligo.org/LIGO-T1500293/public}}, and add the 50\% credible regions of three BNS candidates in O3 to the pattern, including S190425z (orange contours), S190901ap (green contours), and S190910h (purple contours). It implies that candidate S190425z and S190901ap are roughly located in the red region, for which the SNR of the L1 detector is higher than that of H1, and the absence of the H1 data is expected to not effectively influence the estimate of $\tilde{\Lambda}$. As for the candidate S190910h, the situation is unclear because of the very poor localization due to the sole detection by the L1 detector. Note that all these three candidates took place considerably more distant (S190425z, S190901ap, and S190910h are at the distances of $156\pm41$ Mpc, $241\pm79$ Mpc, and $230\pm88$ Mpc, respectively) than that of GW170817 and the SNRs are expected to be lower, which may render the estimated $\tilde{\Lambda}$ biased to larger values. But such an interesting tendency is likely outshone by the uncertainties of the estimation in the low SNR cases.

%%%%%%%%%%%%%%%%%%%%%%%%%%%%%%%%% Conclusion %%%%%%%%%%%%%%%%%%%%%%%%%%%%%%%
\section{Summary}
\label{sec:Summary}
In this work, we firstly adopt the Bayesian inference to re-analyze the individual detector data of GW170817 and resolve the roles of the L1 and H1 data on constraining the properties of the neutron stars. The H1 data favor a compact NS, while the L1 data allow for a more extended NS (see Fig.\ref{fig:lam12}). Then we use the parameterized EoS method to infer the posterior distributions of tidal parameter and radius at $M=1.4M_{\odot}$, namely $\Lambda_{1.4}$ and $R_{1.4}$. The behaviors of the distributions of $\Lambda_{1.4}$ and $R_{1.4}$ are rather similar. For each individual detector, the probability distributions of $\Lambda_{1.4}$ and $R_{1.4}$ have just a single peak, while the combined HLV data yield bimodal distribution (see Fig.\ref{fig:LR}). Our results basically confirm that of \cite{2018arXiv181206100N}, where the authors calculated $\tilde{\Lambda}$.

To better reveal the underlying reasons for the different performance of L1 and H1 detectors on bounding the $\Lambda$ we carried out some simulations. We make some injections with different SNRs by varying different distance, and get the relation between the SNR and the parameter estimation result of $\tilde{\Lambda}$. One general tendency found in our simulation results is the lower the SNR, the higher the $\tilde{\Lambda}$. Furthermore, we adopt the method developed by \cite{2012PhRvD..85l3007D} to compute the measurabilities of the chirp mass and tidal deformability. It turns out that the measurabilities of $\mathcal{M}$ and $\Lambda$ are mainly governed by the low frequency ($<100$ Hz) and high frequency ($>100$ Hz) data, respectively. Together with the analysis of different low-frequency-cutoff data of GW170817, we show that the L1 data has a SNR lower than that of H1 at high frequencies. This seems helpful in explaining the fact that the L1 data favor a higher $\tilde{\Lambda}$, $\Lambda_{1.4}$ and $R_{1.4}$ than those of the H1 data. However, the SNR of H1 is just slightly higher than L1 at frequencies above $\sim 100$ Hz, which is insufficient to account for the difference showing in the data analysis results. In fact, the different probability distributions of the tidal parameters favored by the L1 and H1 data of GW170817 are reasonable in view of the large fluctuation of the expected tidal parameter in the cases of low SNR (see Fig.\ref{fig:ld_dis}).

Finally, we study the effect of sky location on the SNRs of the two detectors, and the result shows a pattern for the difference between the two detectors. We then use O3's PSDs to make a ``map" for the undergoing third observing run, where we can establish a relation between the sky location and the sensitivity for tidal deformability parameter of the two Advanced LIGO detectors. We add the 50\% credible regions of the localization of the BNS candidates S190425z, S190901ap, and S190910h to the ``map". S190425z and S190901ap roughly located in the regions favoring the L1 detector, and the absence of the H1 data is expected to not effectively influence the estimation of $\tilde{\Lambda}$. As for the candidate S190910h, the situation is unclear because of the very poor localization due to the sole detection by the L1 detector. More importantly, all these three candidates took place significantly more distant than that of GW170817 and the SNRs are expected to be lower, which may render the estimated $\tilde{\Lambda}$ biased to a higher value. Such an interesting tendency however is expected to be disguised by the uncertainties of the parameter estimation in the low SNR cases.

\section*{ACKNOWLEDGMENTS}
This work was supported in part by NSFC under grants of No. 11525313 (i.e., Funds for Distinguished Young Scholars), No. 11921003 and No. 11433009. This research has made use of data and software obtained from the Gravitational Wave Open Science Center \url{https://www.gw-openscience.org}, a service of LIGO Laboratory, the LIGO Scientific Collaboration and the Virgo Collaboration. LIGO is funded by the U.S. National Science Foundation. Virgo is funded by the French Centre National de Recherche Scientifique (CNRS), the Italian Istituto Nazionale della Fisica Nucleare (INFN) and the Dutch Nikhef, with contributions by Polish and Hungarian institutes.

\software{PyCBC \citep[version 1.13.6, ascl:1805.030, \url{http://doi.org/10.5281/zenodo.3265452}]{2018ascl.soft05030T}, Bilby \citep[version 0.5.5, ascl:1901.011, \url{https://git.ligo.org/lscsoft/bilby/}]{2019ascl.soft01011A}, PyMultiNest \citep[version 2.6, ascl:1606.005, \url{https://github.com/JohannesBuchner/PyMultiNest}]{2016ascl.soft06005B}, LALSuite \citep[version 6.57, \url{https://doi.org/10.7935/GT1W-FZ16}]{lalsuite}}

\end{document}